\documentclass[prl,floatfix,showpacs,amsmath,twocolumn]{revtex4}

\usepackage{amssymb}
\usepackage[mathscr]{eucal}
\usepackage[dvips]{graphicx}

\DeclareMathOperator{\X}{X}
\DeclareMathOperator{\XX}{\mathbb{X}}
\DeclareMathOperator{\Xt}{\tilde{X}}
\DeclareMathOperator{\Xb}{\bar{X}}
\DeclareMathOperator{\XXt}{\tilde{\mathbb{X}}}
\DeclareMathOperator{\OO}{O}
\DeclareMathOperator{\tr}{Tr}

\begin{document}

\newcommand{\mywidth}{0.3\textwidth}
\newcommand{\spec}{\textrm{spec}}
\newcommand{\Span}{\textrm{span}}
\newcommand{\ket}[1]{| \, #1 \, \rangle}
\newcommand{\bra}[1]{\langle \, #1 \, |}
\newcommand{\proj}[1]{\ket{#1} \bra{#1}}
\newcommand{\Proj}[2]{\ket{#1}_{#2} \bra{#1}}
\newcommand{\scal}[2]{\bra{#1} \, #2 \, \rangle}
\newcommand{\expect}[1]{\langle #1 \rangle}
\newcommand{\Expect}[2]{\langle #1 \rangle_{#2}}
\newcommand{\HH}{\mathcal{H}}
\newcommand{\adj}[1]{#1^{\dagger}}
\newcommand{\binomial}[2]{{#1 \choose #2}}
\newcommand{\round}[1]{\left[ #1 \right]}
\newcommand{\roundceil}[1]{\left\lceil #1 \right\rceil}
\newcommand{\roundfloor}[1]{\left\lfloor #1 \right\rfloor}
\newcommand{\Varepsilon}{\mathscr{E}}
\newcommand{\EE}{\mathcal{E}}
\newcommand{\arccot}{\textrm{arccot}}
\newcommand{\sgn}{\textrm{sgn}}
\newcommand{\ketr}{| \rightarrow \rangle}
\newcommand{\ketl}{| \leftarrow \rangle}
\newcommand{\ketu}{| \uparrow \, \rangle}
\newcommand{\ketd}{| \downarrow \, \rangle}
\newcommand{\brar}{\langle \rightarrow |}
\newcommand{\sx}[1]{\sigma_x^{(#1)}}
\newcommand{\sy}[1]{\sigma_y^{(#1)}}
\newcommand{\sz}[1]{\sigma_z^{(#1)}}
\newcommand{\splus}[1]{\sigma_{#1}^{+}}
\newcommand{\sminus}[1]{\sigma_{#1}^{-}}
\newcommand{\fpi}[2]{\left( \frac{2 \pi #1}{#2} \right) }
\newcommand{\omicron}[1]{\textrm{O} \left[ \frac{1}{#1} \right] }
\newcommand{\Omicron}[1]{\textrm{O} \left[ #1 \right] }
\newcommand{\order}[1]{\left( \frac{1}{#1} \right)}
\newcommand{\Xp}[2]{\sideset{^{#1}}{_{#2}}\X}
\newcommand{\XXp}[2]{\sideset{^{#1}}{_{#2}}\XX}
\newcommand{\Xpt}[2]{\sideset{^{#1}}{_{#2}}\Xt}
\newcommand{\Xpb}[2]{\sideset{^{#1}}{_{#2}}\Xb}
\newcommand{\Xppt}[3]{\sideset{^{#1}}{_{#2}^{#3}}\Xt}
\newcommand{\XXpt}[2]{\sideset{^{#1}}{_{#2}}\XXt}
\newcommand{\Op}[2]{\sideset{^{#1}}{_{#2}}\OO}

\title{Efficient evaluation of partition functions of frustrated and inhomogeneous spin systems}

\author{V. Murg$^1$, F. Verstraete$^{1,2}$, J. I. Cirac$^1$}
\affiliation{$^1$Max-Planck-Institut f\"ur Quantenoptik, Hans-Kopfermann-Str. 1,
Garching, D-85748, Germany\\
$^2$Institute for Quantum Information, Caltech, Pasadena, US}

\pacs{03.67.-a, 03.75.Lm, 02.70.-c, 75.40.Mg}
\date{\today}

\begin{abstract}
We present a numerical method to evaluate partition functions and
associated correlation functions of inhomogeneous 2--D classical
spin systems and 1--D quantum spin systems. The method is scalable
and has a controlled error. We illustrate the algorithm by
calculating the finite--temperature properties of bosonic
particles in 1--D optical lattices, as realized in current
experiments.
\end{abstract}

\maketitle


The study of thermodynamic properties of 2--D classical spin
systems has a very long and fruitful history. On the one hand, it
has provided valuable insights in the theory of magnetism and
phase transitions. On the other, it has allowed us to describe
1--D quantum systems at finite temperature by virtue of the
Suzuki--Trotter decomposition.

As very few spin models are exactly solvable, many different
approximate methods have been proposed to calculate the associated
partition functions. Monte Carlo seems to be the method of choice
for non--frustrated systems, but fails in the description of
frustrated and fermionic systems due to the notorious sign problem
\cite{troyer04}. No such sign problem occurs in the method
developed by Nishino~\cite{nishino95}, where the largest
eigenvalue of the transfer matrix of the classical spin model can
be approximated by using a variation of the density matrix
renormalization group (DMRG) approach
\cite{white92,schollwoeck04}. By making use of the Suzuki--Trotter
decomposition, this method has also been used to calculate the
free energy of translational invariant 1-D quantum systems
\cite{bursill96,shibata97,wang97}. The main restriction of this
method is that it cannot be applied in situations in which the
number of particles is finite and/or the system is not
homogeneous. The method may also become ill conditioned when the
transfer matrix is not hermitian. Finally, in the case of 1-D
quantum systems, a recent development
\cite{verstraeteripoll04,zwolak04} allows us to overcome these
problems by extending the concept of matrix product states
\cite{fannes92,verstraetedelgado04} to mixed states, e.g. by using
the idea of purification of states \cite{verstraeteripoll04}. This
method is, however, specially designed for 1--D quantum systems,
and cannot be extended to classical 2--D models.

Here we take a completely different approach which allows us to
overcome the drawbacks of the above mentioned methods in both 2--D
classical and 1--D quantum systems. We achieve this by evaluating
the associated partition and correlation functions directly. The
main advantages of this method are that the approximations made
are very well controlled, that it applies to frustrated,
inhomogeneous and finite classical and quantum systems, and that
it can be generalized to higher dimensions. We will illustrate its
performance with the system of strongly interacting bosonic atoms
in optical lattices, a problem which has attracted a lot of
interest in the atomic physics community in the last few years due
to the recent experimental
achievements~\cite{stoeferle04,paredes04,koehl04}. In this system,
20-100 atoms are trapped by the combination of a periodic and a
harmonic (i.e. inhomogeneous) potential created by lasers in 1--D
and at a finite temperature. In the so--called Tonks--Girardeau
limit~\cite{girardeau60}, the problem can be exactly solved via
fermionization~\cite{paredes04}, and thus this provides us with a
reliable benchmark for our method. Outside this limit, we are able
to reproduce certain features experimentally
observed~\cite{stoeferle04}.

Our method relies in reexpressing the partition and correlation
functions as a contraction of a collection of 4-index tensors,
which are disposed according to a 2--D configuration. We will
perform this task for both 2--D classical and 1--D quantum
systems. We will then show how the methods introduced
in~\cite{verstraetecirac04} can be used to approximate these
contractions in a controlled way, and thus lead to a scalable
algorithm for the evaluation of the quantities of interest.


{\em 1.- 2--D classical systems:} Let us consider first the
partition function of an inhomogeneous classical 2--D n--level
spin system on a $L_1\times L_2$ lattice. For simplicity we will
concentrate on a square and nearest--neighbor interactions,
although our method can be easily extended to other short--range
situations. We have
\begin{displaymath}
    Z= \sum_{x^{1 1},\ldots,x^{L_1 L_2}}\exp\left[-\beta H(x^{1 1},\ldots,x^{L_1 L_2}) \right],
\end{displaymath}
where
\begin{displaymath}
    H\left(x^{1 1},\ldots\right) = \sum_{i j} \left[ H_{\downarrow}^{ij}\left(x^{i j}, x^{i+1,j}\right) +
    H_{\rightarrow}^{i j}\left(x^{i j}, x^{i,j+1}\right) \right]
\end{displaymath}
is the Hamiltonian, $x^{i j}=1,\ldots,n$ and $\beta$ is the
inverse temperature. The singular value decomposition allows us to
write
\begin{displaymath}
    \exp\left[ -\beta H_{q}^{ij}(x,y)\right]=\sum_{\alpha=1}^{n}
    f_{q\alpha}^{ij}(x)g_{q\alpha}^{ij}(y),
\end{displaymath}
with $q \in \{\downarrow, \rightarrow \}$. Defining the tensors
\begin{displaymath}
    X_{lrud}^{ij}=\sum_{x=1}^n
    f_{\downarrow d}^{ij}(x)g_{\downarrow u}^{ i-1,j}(x)f_{\rightarrow r}^{ij}(x)g_{\rightarrow l}^{i,j-1}(x),
\end{displaymath}
the partition function can now be calculated by contracting all
4-index tensors $X^{ij}$ arranged on a square lattice in such a
way that, e.g., the indices $l,r,u,d$ of $X^{ij}$ are contracted
with the indices $r,l,d,u$ of the respective tensors
$X^{i,j-1},X^{i,j+1},X^{i-1,j},X^{i+1,j}$. In order to determine
the expectation value of a general operator of the form $O(\{x^{i
j}\})=Z \prod_{i j} O^{i j}(x^{i j})$, one just has to replace
each tensor $X^{i j}$ by
\begin{displaymath}
     X_{lrud}^{i j}\left(O^{i j}\right)=\sum_{x=1}^n O^{i j}(x)
     f_{\downarrow d}^{ij}(x) g_{\downarrow u}^{ i-1,j}(x) f_{\rightarrow r}^{ij}(x) g_{\rightarrow l}^{i,j-1}.
\end{displaymath}


{\em 2.- 1--D quantum systems:} We consider the partition function
of an inhomogeneous 1--D quantum system composed of $L$ $n$-level
systems,
\begin{displaymath}
    Z=\tr \exp\left(-\beta H\right).
\end{displaymath}
It is always possible to write the Hamiltonian $H$ as a sum
$H=\sum_k H_k$ with each part consisting of a sum of commuting
terms. Let us, for simplicity, assume that $H=H_1+H_2$ and that
only local and 2-body nearest neighbor interactions occur, i.e.
$H_k=\sum_i O^{i,i+1}_k$ and
$\left[O^{i,i+1}_k,O^{j,j+1}_k\right]=0$, with $i,j=1,\ldots,L$.
The more general case can be treated in a similar way. Let us now
consider a decomposition
\begin{equation} \label{eqn:decompop}
     \exp\left(-\frac{\beta}{M} O^{i,i+1}_k\right)=\sum_{\alpha=1}^{\kappa}
     \hat{S}^{i}_{k \alpha}\otimes
     \hat{T}^{i+1}_{k \alpha}.
\end{equation}
The singular value decomposition guarantees the existence of such
an expression with $\kappa \leq n^2$. As we will see later, a
smart choice of $H=\sum_k H_k$ can typically decrease~$\kappa$
drastically. Making use of the Suzuki--Trotter
formula~\footnote{Note that in practice, it will be desirable to
use the higher order versions of the Trotter decomposition.}
\begin{displaymath}
     Z={\rm Tr} \left(\prod_k\exp\left(-\frac{\beta}{M} H_k\right)\right)^M
     + \Omicron{\frac{1}{M}}
\end{displaymath}
it can be readily seen that the partition function can again be
calculated by contracting a collection of 4-index tensors $X^{ij}$
defined as
\begin{displaymath}
    X^{ij}_{(ll')(rr')ud} \equiv
    \left[ \hat{T}^{j}_{1 l} \hat{S}^{j}_{1 r} \hat{T}^{j}_{2 l'} \hat{S}^{j}_{2 r'}
    \right]_{\left[u d\right]},
\end{displaymath}
where the indices $(l,l')$ and $(r,r')$ are combined to yield a
single index that may assume values ranging from~$1$
to~$\kappa^2$. Note that now the tensors $X^{ij}$ and $X^{i'j}$
coincide, and that the indices~$u$ of the first and~$d$ of the
last row have to be contracted with each other as well, which
corresponds to a classical spin system with periodic boundary
conditions in the vertical direction. A general expectation value
of an operator of the form $O=Z O^1 \otimes \cdots \otimes O^N$
can also be reexpressed as a contraction of tensors with the same
structure: it is merely required to replace each tensor~$X^{1 j}$
in the first row by
\begin{displaymath}
    X^{1 j}_{(ll')(rr')ud} \left( O^j \right) =
    \left[ O^j \hat{T}^{j}_{1 l} \hat{S}^{j}_{1 r} \hat{T}^{j}_{2 l'} \hat{S}^{j}_{2 r'}
    \right]_{\left[u d\right]}.
\end{displaymath}


{\em 3.- Tensor contraction:} In the following, we adapt the
algorithm introduced in \cite{verstraeteporras04} in order to
contract the tensors $X^{ij}$ introduced above in a controlled
way. The main idea is to express the objects resulting from the
contraction of tensors along the first and last column in the 2--D
configuration as matrix product states (MPS) and those obtained
along the columns $2,3,\ldots,L-1$ as matrix product operators
(MPO)~\cite{verstraeteripoll04}. More precisely, we define
 \begin{eqnarray*}
 \bra{\XX^1} & := & \sum_{r_1 \ldots r_M=1}^{m}
 \tr \left( \X^{1 1}_{r_1} \ldots \X^{M 1}_{r_M} \right)
 \bra{r_1 \ldots r_M}\\
 \ket{\XX^L} & := & \sum_{l_1 \ldots l_M=1}^{m}
 \tr \left( \X^{1 L}_{l_1} \ldots \X^{M L}_{l_M} \right)
 \ket{l_1 \ldots l_M}\\
 \XX^j & := & \sum_{l_1,r_1,\ldots=1}^{m}
 \tr \left( \X^{1 j}_{l_1 r_1} \ldots \X^{M j}_{l_M r_M} \right)
 \ket{l_1 \ldots} \bra{r_1 \ldots},
 \end{eqnarray*}
where $m=n$ for 2--D~classical systems and $m=\kappa^2$ for
1--D~quantum systems. These MPS and MPOs are associated to a chain
of $M$ $m$--dimensional systems and their virtual dimension
amounts to $D=n$. Note that for 2--D~classical systems the first
and last matrices under the trace in the MPS and MPO reduce to
vectors. The partition function (and similarly other correlation
functions) reads $Z=\bra{\XX^1} \XX^{2} \cdots \XX^{L-1}
\ket{\XX^{L}}$.
Evaluating this expression iteratively by calculating step by step
$\bra{\XX^j}:=\bra{\XX^{j-1}} \XX^{j}$ for $j=2,\ldots,L-1$ fails
because the virtual dimension of the MPS $\bra{\XX^{j}}$ increases
exponentially with~$j$. A way to circumvent this problem is to
replace in each iterative step the MPS $\bra{\XX^{j}{}}$ by a MPS
$\bra{\XXt^{j}}$ with a reduced virtual dimension~$\tilde{D}$ that
approximates the state $\bra{\XX^{j}}$ best in the sense that the
norm $\delta K:=\|\bra{\XX^{j}}-\bra{\XXt^{j}} \|$ is minimized.
Due to the fact that this cost function is multiquadratic in the
variables of the MPS, this minimization can be carried out very
efficiently~\cite{verstraetecirac04,verstraeteporras04,verstraeteripoll04};
the exponential increase of the virtual dimension can hence be
prevented and the iterative evaluation of~$Z$ becomes tractable,
such that an approximation to the partition function can be
obtained from $Z \simeq \scal{\XXt^{L-1}}{\XX^{L}}$. The accuracy
of this approximation depends only on the choice of the reduced
dimension~$\tilde{D}$ and the approximation becomes exact
for~$\tilde{D} \geq D^L$. As the norm $\delta K$ can be calculated
at each step, $\tilde{D}$ can be increased dynamically if the
obtained accuracy is not large enough. In the worst case scenario,
such as in the NP-complete Ising spin glasses~\cite{barahona82},
$\tilde{D}$ will probably have to grow exponentially in $L$ for a
fixed precision of the partition function. But in less
pathological cases it seems that $\tilde{D}$ only has to grow
polynomially in $L$; indeed, the success of the methods developed
by Nishino~\cite{nishino95} in the translational invariant case
indicate that even a constant $\tilde{D}$ will produce very
reliable results.


{\em 4.- Illustration: Bosons in optical lattices:} A system of
trapped bosonic particles in a 1--D optical lattice of~$L$ sites
is described by the Bose-Hubbard Hamiltonian~\cite{jaksch98}
\begin{displaymath}
 H = -J \sum_{i=1}^{L-1} ( \adj{a}_i a_{i+1} + h.c.) + \frac{U}{2}
 \sum_{i=1}^L  \hat{n}_i (\hat{n}_i-1) +
 \sum_{i=1}^L V_i \hat{n}_i,
\end{displaymath}
where $\adj{a}_i$ and $a_i$ are the creation and annihilation
operators on site~$i$ and $\hat{n}_i=\adj{a}_i a_i$ is the number
operator. This Hamiltonian describes the interplay between the
kinetic energy due to the next-neighbor hopping with amplitude~$J$
and the repulsive on-site interaction~$U$ of the particles. The
last term in the Hamiltonian models the harmonic confinement of
magnitude $V_i = V_0 (i-i_0)^2$. The variation of the ratio~$U/J$
drives a phase-transition between the Mott-insulating and the
superfluid phase, characterized by localized and delocalized
particles respectively~\cite{fisher89}. Experimentally, the
variation of~$U/J$ can be realized by tuning the depth of the
optical lattice~\cite{jaksch98,buechler03}. On the other hand, one
typically measures directly the momentum distribution by letting
the atomic gas expand and then measuring the density distribution.
Thus, we will be mainly interested here in the (quasi)--momentum
distribution
 \begin{displaymath}
 n_k = \frac{1}{L} \sum_{r,s=1}^{L} \expect{\adj{a}_r a_s} e^{i 2 \pi k
 (r-s)/L}.
 \end{displaymath}

Our goal is now to study with our numerical method the
finite-temperature properties of this system for different
ratios~$U/J$. We thereby assume that the system is in a thermal
state corresponding to a grand canonical ensemble with chemical
potential~$\mu$, such that the partition function is obtained as
$Z=\tr e^{-\beta(H - \mu \hat{N})}$. Here, $\hat{N}=\sum_{i=1}^L
\hat{n}_i$ represents the total number of particles. For the
numerical study, we assume a maximal particle--number~$q$ per
lattice site, such that we can project the Hamiltonian~$H$ on the
subspace spanned by Fock-states with particle-numbers per site
ranging from~$0$ to~$q$. The projected Hamiltonian~$\tilde{H}$
then describes a chain of~$L$ spins, with each spin acting on a
Hilbert-space of dimension~$n=q+1$. A Trotter decomposition that
turned out to be advantageous for this case is
\begin{equation} \label{eqn:decomp_bosehubbard}
    e^{-\beta (\tilde{H} - \mu \hat{N} )}= \left( \adj{\hat{V}} \hat{V} \right)^M+\Omicron{\frac{1}{M^2}},
\end{equation}
with $\tilde{H}=H_R+H_S+H_T$, $H_R = - \frac{J}{2}
\sum_{i=1}^{L-1} R^{(i)} R^{(i+1)}$, $H_S = - \frac{J}{2}
\sum_{i=1}^{L-1} S^{(i)} S^{(i+1)}$, $H_T = \sum_{i=1}^L T^{(i)}$,
$R^{(i)}=\adj{\tilde{a}}_i+\tilde{a}_i$, $S^{(i)}=-i
(\adj{\tilde{a}}_i-\tilde{a}_i)$, $T^{(i)}=\frac{1}{2} \tilde{n}_i
(\tilde{n}_i-1)+V_i \tilde{n}_i$ and $\hat{V} = e^{-\frac{\beta}{2
M} H_R} e^{-\frac{\beta}{2M} H_S} e^{-\frac{\beta}{2M} (H_T-\mu
\hat{N})}$. $\adj{\tilde{a}}_i$, $\tilde{a}_i$ and $\tilde{n}_i$
thereby denote the projections of the creation, the annihilation
and the number operators $\adj{a}_i$, $a_i$ and $n_i$ on the
$q$-particle subspace. The decomposition~(\ref{eqn:decompop}) of
all two-particle operators in
expression~(\ref{eqn:decomp_bosehubbard}) then straightforwardly
leads to a set of 4-index tensors $X^{i j}_{l r u d}$, with
indices~$l$ and~$r$ ranging from~$1$ to~$(q+1)^3$ and indices~$u$
and~$d$ ranging from~$1$ to~$q+1$. Note that the typical second
order Trotter decomposition with $H=H_{\rm even}+H_{\rm odd}$
would make the indices~$l$ and~$r$ range from~$1$ to~$(q+1)^6$.

\begin{figure}[t]
    \begin{center}
        \includegraphics[width=0.49\textwidth]{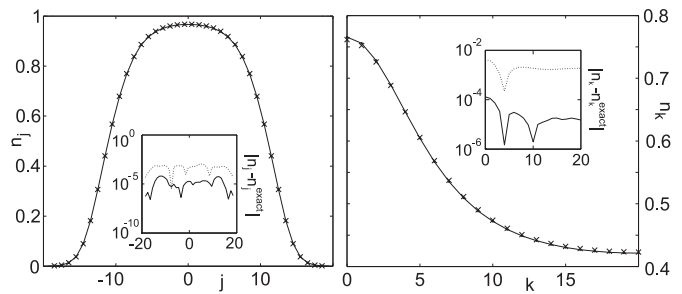}
    \end{center}
    \caption{
        Density and (quasi)-momentum distribution in the Tonks-Girardeau gas limit, plotted for $\beta J=1$,
        $L=40$, $N=21$ and $V_0/J=0.034$. The dots (crosses) represent the numerical
        results for $\tilde{D}=2$ ($\tilde{D}=8$) and the solid
        line illustrates the exact results. From the
        insets, the error of the numerical
        results can be gathered.
        }
    \label{fig:tonks}
\end{figure}

Let us start out by considering the limit $U/J \to \infty$ in
which double occupation of single lattice sites is prevented and
the particles in the lattice form a Tonks--Girardeau
gas~\cite{paredes04}. In this limit, the Bose-Hubbard Hamiltonian
maps to the Hamiltonian of the exactly solvable (inhomogeneous)
XX-model, which allows to benchmark our algorithm. The comparison
of our numerical results to the exact results can be gathered from
fig.~\ref{fig:tonks}. Here, the density and the (quasi)-momentum
distribution are considered for the special case $\beta J=1$,
$L=40$, $N=21$ and $V_0/J=0.034$. The numerical results shown have
been obtained for Trotter-number~$M=10$ and two different reduced
virtual dimensions~$\tilde{D}=2$ and~$\tilde{D}=8$. The
norm~$\delta K$ was of order~$10^{-4}$ for $\tilde{D}=2$
and~$10^{-6}$ for~$\tilde{D}=8$~\footnote{We note that we have
stopped our iterative algorithm at the point the variation
of~$\delta K$ was less than~$10^{-8}$.}. From the insets, it can
be gathered that the error of the numerical calculations is
already very small for~$\tilde{D}=2$ (of order $10^{-3}$) and
decreases significantly for~$\tilde{D}=8$. This error can be
decreased further by increasing the Trotter-number~$M$.

\begin{figure}[t]
    \begin{center}
        \includegraphics[width=0.45\textwidth]{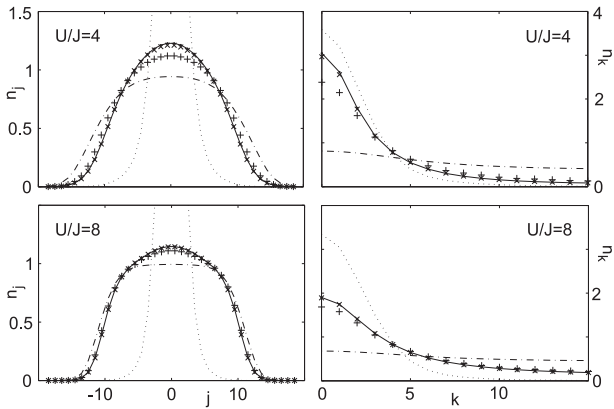}
    \end{center}
    \caption{
        Density and (quasi)-momentum distributions for interaction strengths
        $U/J=4$ and~$8$. Here, $\beta J=1$,
        $L=40$, $N=21$ and $M=10$.
        Numerical results were obtained for~$q=2$ (plus-signs), $q=3$ (crosses) and~$q=4$ (solid line).
        For comparison, the distributions for
        $U/J=0$ (dotted lines) and $U/J \to \infty$ (dash-dotted lines) are also
        included.
        }
    \label{fig:bosehubbard_mdist}
\end{figure}

As the ratio~$U/J$ becomes finite, the system becomes physically
more interesting, but lacks an exact mathematical solution. In
order to judge the reliability of our numerical solutions in this
case, we check the convergence with respect to the free parameters
of our algorithm ($q$, $\tilde{D}$ and $M$). As an illustration,
the convergence with respect to the parameter~$q$ is shown in
figure~\ref{fig:bosehubbard_mdist}. In this figure, the density
and the (quasi)-momentum distribution are plotted for $q=2,3$
and~$4$. We thereby assume that $\beta J=1$, $L=40$ and $N=21$ and
consider interaction strengths~$U/J=4$ and~$8$. The harmonic
potential~$V_0$ is chosen in a way to describe Rb-atoms in a
harmonic trap of frequency~$\textrm{Hz}$ (along the lines
of~\cite{paredes04}). We note that we have taken into account that
changes of the ratio~$U/J$ are obtained from changes in both the
on-site interaction~$U$ and the hopping amplitude~$J$ due to
variations of the depth of the optical lattice. The numerical
calculations have been performed with~$M=10$ and $\tilde{D}=q+1$.
From the figure it can be gathered that convergence with respect
to~$q$ is achieved for $q\geq3$.

\begin{figure}[t]
    \begin{center}
        \includegraphics[width=0.45\textwidth]{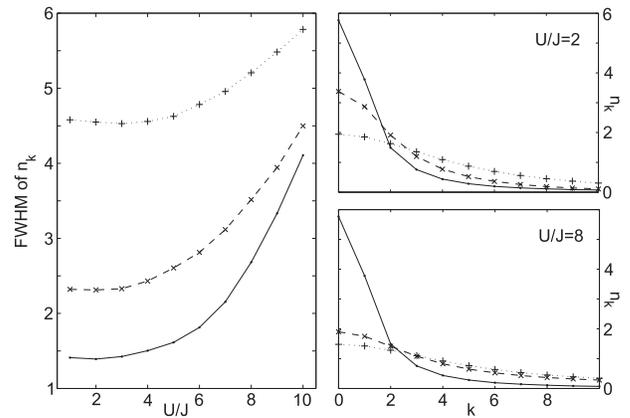}
    \end{center}
    \caption{
        FWHM of the (quasi)-momentum distribution as a function of~$U/J$, calculated for temperatures $\beta J=0.5$
        (plus-sign),$\beta J=1$ (crosses) and $\beta J=2$ (dots).
        The corresponding (quasi)-momentum distributions
        for $U/J=2$ and $U/J=8$ are illustrated in the plots at the right-hand side.
        }
    \label{fig:bosehubbard_FWHM}
\end{figure}

We now use our numerical algorithm to study a physical property of
interacting bosons in an optical lattice, namely the full width at
half maximum (FWHM) of the (quasi)-momentum distribution. It has
been predicted that the FWHM shows a kink at zero
temperature~\cite{kollath04,wessel04,pollet04}. This kink is an
indication for a Mott-superfluid transition, since the FWHM is
directly related to the inverse correlation length. Experiments
have also revealed this kink~\cite{koehl04,stoeferle04}; they are,
however, performed at finite temperature, something we can study
with our algorithm. In figure~\ref{fig:bosehubbard_FWHM}, we plot
the numerical results for the FWHM as a function of~$U/J$ for
three different (inverse) temperatures~$\beta J=0.5$,$1$ and~$2$.
The physical parameters $L$, $N$ and $V_0$ are thereby chosen as
in the previous case. The numerical results have been obtained for
$M=10$, $q=4$ and $\tilde{D}=q+1$. For each temperature, three
different regions can be distinguished: the superfluid region with
constant FWHM, the Mott-region with linearly increasing FWHM and
an intermediate region in which both phases coexist. The
value~$U/J$ at which the Mott-region starts increases with
increasing temperature, which is consistent with the criteria $U
\gg k_B T, J$ for the appearance of the Mott-phase. This behaviour
could be easily observed in present experiments.

In summary, we have presented a numerical method for the
investigation of thermal states of inhomogeneous 2--D classical
and 1--D quantum systems. We have illustrated the usefulness of
this method by applying it to a system of trapped bosonic
particles in a 1--D optical lattice -- which is of current
experimental interest. For this system, we have studied the error
and the convergence of the method. In addition, we have used the
method to study some physical properties of this system and we
have obtained results which can be verified experimentally.


We thank M. A. Martin-Delgado and D. Porras for discussions. Work
supported by the DFG, european projects (IST and RTN) and the
Kompetenznetzwerk der Bayrischen Staatsregierung
Quanteninformation.


\end{document}